\documentstyle[preprint,prl,aps,epsf]{revtex}
\begin{document}
\tightenlines

\title{Pbar Annihilation in $Au+Au$ at AGS Energies}  

\author{Y.~Pang$^{1,3}$, D.~E.~Kahana$^{2}$, S.~H.~Kahana$^{1}$, H.~Crawford$^{4}$}
\address{$^{1}$Physics Department, Brookhaven National Laboratory, Upton, NY 11973, USA\\
   $^{2}$Physics Department, State University of New York, Stony Brook, NY 11791, USA\\
   $^{3}$Physics Department, Columbia University, New York, NY 10027, USA\\
   $^{4}$Physics Department, Space Sciences Laboratory, University of 
   California, Berkeley, CA 9472, USA}
\date{\today}  

\maketitle  
  


\begin{figure}
\vbox{\hbox to\hsize{\hfil
 \epsfxsize=6.1truein\epsffile[24 60 577 736]{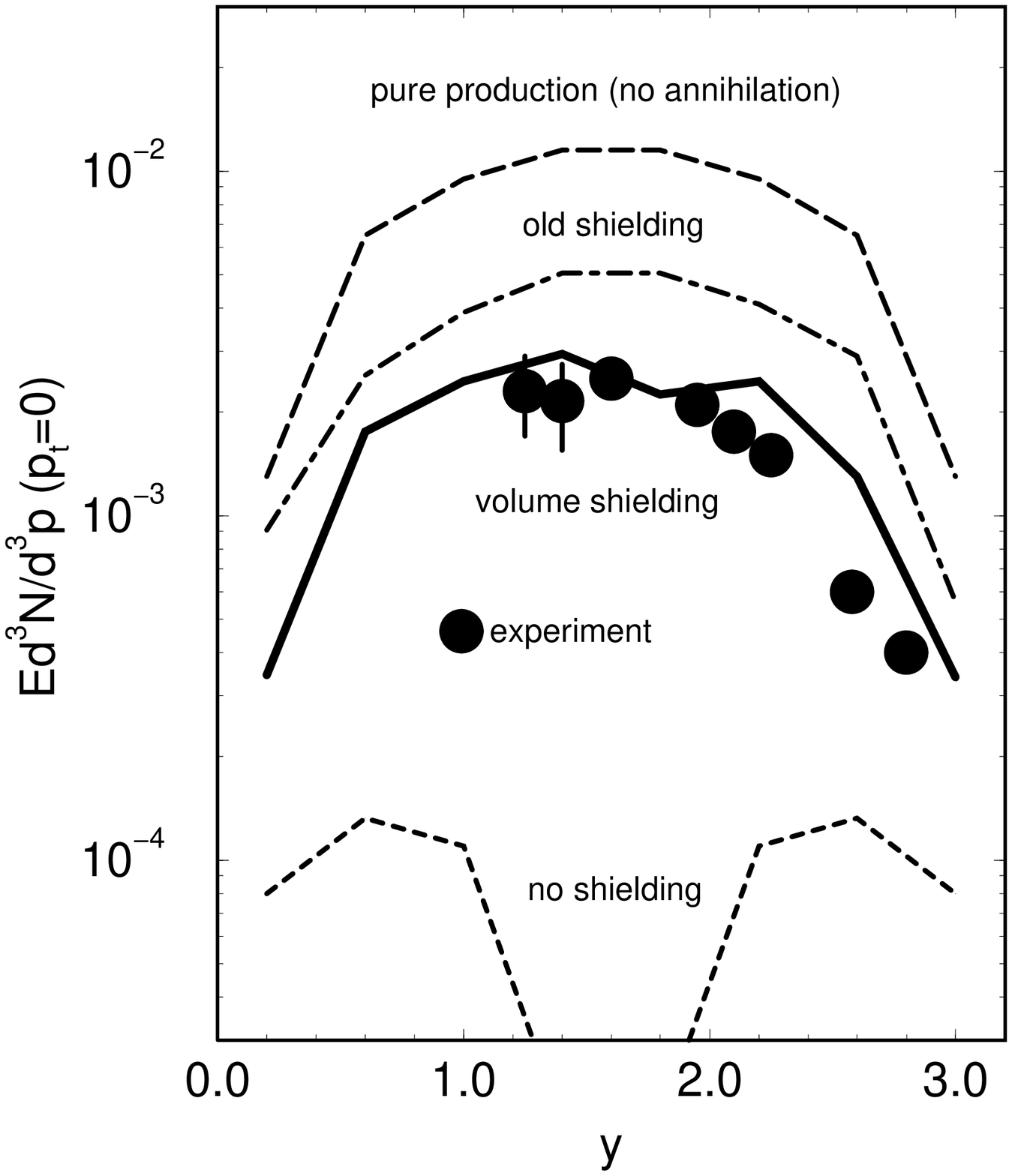}
 \hfil}}
\caption[]{Antiproton production for Au+Au at $11.0\,\,GeV/c$}
\label{fig:one}
\end{figure}

A recent publication from E878 \cite{Crawford} 
reported antiproton yields in a limited
kinematical region near $p_t=0$, for $11\, GeV/c$ $Au+Au$ beams at the
AGS. The data was compared to calculations performed with the cascades RQMD
and ARC and certain conclusions were drawn based on this comparison. What was
missing from this analysis was reference to the overwhelming role that would
have been played by unbridled annihilation, and consequently to the
surprising result that theory is at all close to measurement. This is
apparent from Figure~\ref{fig:one} where we have reproduced the most central
cut from \cite{Crawford} but added some ARC calculations, unfortunately not
included in the latter publication.  First, we call attention to the large
normalisation factor, $\sim 30$ integrating over all $y$ and
$p_t$, between antiproton yields for pure production (but with 
$\bar p$ rescattering) and that for production + unadulterated annihilation,
i.e. without the screening introduced for $Si+Au$ \cite{arc}.

From these data and the previous analysis \cite{arc}, it is apparent that
screening is essential.  This modification in the $p\bar p$ interaction arose
from an insistence on causality. The $p$ and $\bar p$ should not annihilate
before they touch, while the large free-space cross sections allow
annihilation at separations in excess of $3\, fm$. In the medium, other
particles will frequently interfere before any contact, thus generating the
classical shielding. For Si+Au \cite{arc} we introduced a time delay $\tau_a$
calculated dynamically from the $N\bar N$ separation and relative velocity,
permitting annihilation only after {\it complete} overlap. In
Figure~\ref{fig:one} we present Au+Au $\bar p$ yields for an annihliation
probability taken proportional to the overlap volume (volume shielding), as
well as the correct simulations from the old complete overlap \cite{Crawford}.

The results in Figure~\ref{fig:one} are reasonable, fortuitous to some extent
considering the large annihilation factors, but built on a straightforward
modeling. The $Au+Au$ $\bar p$ production rapidity shape is well described by
ARC + screening and the magnitudes can even be approached with some
quantitative understanding.  We took $b<5\,fm$ to closely approximate the
trigger used in Ref.~1.  The variance between ``old'' screening and new
volume-overlap screening, some $35\%$, could be taken as an extreme limit for
the theoretical uncertainty, but with the new, volume, screening designated as
the the theoretical norm since annihilation can commence before complete 
overlap A prime reason for the relative stability of the modeling is the
screening-induced reduction of the large free annihilation cross-section
to something close to that related to the average separation in the nuclear
environment. The cascade is a powerful tool for handling this three body
process, and once some form of shielding is in place the calculated yields
are stable.

Can one use antiparticles as a signal for unusual behaviour at these low
energies?  Probably not at a level one would like.  We do not here, nor did
we in earlier publications ~\cite{arc}, claim a ``good'' quantitative
understanding of low energy $\bar p$'s, but in fact have done much better
than might have been expected.  At the AGS Au beam momenta one may need
measurements of $\sigma (pp \to \bar p)$ rather than relying just on
extrapolation from higher energies. A kinematically more comprehensive set of
$\bar p$ data, perhaps from the AGS experiments E864 and E866, should allow
one to exploit the successes of the modeling presented here.

This manuscript has been authored under DOE supported research Contract
Nos. DE-FG02-93ER40768, DE--AC02--76CH00016, DE-FG02-92~ER40699, and
DE-FG03-90ER-40571. We would also like to acknowledge support from the
Alfred P. Sloan Foundation (Y.~P.), and from the Alexander von Humboldt
Foundation, Bonn, Germany (S.~H.~K.).



\clearpage
\clearpage

\end{document}